# Merging Dirac electrons and correlation effect in the heterostructured $Bi_2Te_3/Fe_{1+\delta}Te$

Guan Du, Zengyi Du, Xiong Yang, Enyu Wang, Delong Fang, Huan Yang & Hai-Hu Wen[*]

National Laboratory of Solid State Microstructures and Department of Physics, Nanjing University, Nanjing 210093, China

The topological insulator and strong electronic correlation effect are two important subjects in the frontier studies of modern condensed matter physics. A topological insulator exhibits a unique pair of surface conduction bands with the Dirac dispersion albeit the bulk insulating behaviour. These surface states are protected by the topological order, and thus the spin and momentum of these surface electrons are locked together demonstrating the feature of time reversal invariance. On the other hand, the electronic correlation effect becomes the very base of many novel electronic states, such as high temperature superconductivity, giant magnetoresistance etc. Here we report the discovery of merging the two important components: Dirac electrons and the correlation effect in heterostructured $Bi_2Te_3/Fe_{1+\delta}Te$. By measuring the scanning tunneling spectroscopy on $Bi_2Te_3$ thin films (a typical topological insulator) thicker than 6 quintuple layers on top of the $Fe_{1+\delta}Te$ single crystal (a parent phase of the iron based superconductors $FeSe_{1-x}Te_x$), we observed the quantum oscillation of Landau levels of the Dirac electrons and the gapped feature at the Fermi energy due to the correlation effect of $Fe_{1+\delta}Te$. Our observation challenges the ordinary understandings and must demonstrate some unexplored territory concerning the combination of topological insulator and strong correlation effect.

The topological insulator (TI) was predicted by theory as a new matter of quantum phase, and quickly realized by experiments[1-3]. TI has an insulating bulk state and topologically protected metallic surface states (SSs). The band structure of SSs has a nearly linear dispersion forming a Dirac cone near the Fermi level. The electrons of SSs which are called Dirac electrons should obey the massless Dirac equation. The SSs are spin nondegenerate and spin-momentum locked satisfying time reversal symmetry. Many peculiar characteristics of TIs have been studied. However, most of the efforts are devoted to the underlying physics of non-interacting electrons. The question remains how the Dirac electrons behave under the effect of the electron interactions. Although the strong correlated topological Kondo insulator $SmB_6$ has been widely studied[4-6], the narrow bulk gap appears to be the obstacle for the detecting of SSs in that system.

The electronic correlation effect is the very base of many novel electronic states in condensed matter physics. It has been assumed that the parent compounds of cuprate superconductors can be regarded as Mott insulators with strong correlation effect. How the high-$T_c$ superconductivity emerges by doping charges into the insulating compound is still unclear, and it is one of the greatest challenges in modern condensed matter physics[7]. Another kind of high-temperature superconductors, i.e., iron-based superconductors, are also considered to possess correlation effect to some extent[8-11]. $Fe_{1+\delta}Te$, the parent compound of superconducting 11-family[12], is supposed to be a system with the strongest electronic correlation effect in the iron based superconductors[10].

Heterostructures based on TIs and other compounds have led to the discovery of many novel quantum phases, such as possible topological superconductivity induced by proximity effect[13,14]. The recent experiment of magnetic Fe chains on top of the conventional superconductor Pb exhibit the possible evidence of the long time sought Majorana fermions[15]. Constructing heterostructures with van der Waals interactions has already become an effective method for the research of these novel quantum states of TIs. $Bi_2Te_3$ is a typical TI with a large bulk gap of about 140 meV together with linear dispersion surface states[16-18]. $Bi_2Te_3$ thin films can be grown on a variety of substrates thanks to the van der Waals interactions[19-22].

In order to introduce the strong correlation effect into TI and study the behaviours of the Dirac electrons, we grew $Bi_2Te_3$ thin films on the surface of cleaved $Fe_{1+\delta}Te$ single crystal by molecular beam epitaxy (MBE) method. We also applied scanning tunnelling microscopy/spectroscopy (STM/STS) studies both on $Fe_{1+\delta}Te$ and $Bi_2Te_3/Fe_{1+\delta}Te$ heterostructures. A dip or a V-shaped structure on the density of states (DOS) around the Fermi level is observed in the STS measurement on $Fe_{1+\delta}Te$ which is the signature of strong correlation effect. The zero-bias dip structure appears again in the STS spectra measured on the $Bi_2Te_3$ thin film grown on $Fe_{1+\delta}Te$ substrate. We demonstrate that the suppressing of DOS is caused by the strong correlation effect inherited from the $Fe_{1+\delta}Te$ substrate instead of the intrinsic band structure of $Bi_2Te_3$. The Dirac electrons and the strong correlation effect are merged. Considering the thickness of several nanometres of $Bi_2Te_3$ films, how the strong correlation effect is delivered from the bottom to the top surface of the TI film is mysterious and challenges the ordinary understandings. We believe this novel phenomenon must tie closely to the unique feature of the topological SSs.

The $Fe_{1+\delta}Te$ single crystal has layered structure, and the neighbour layers are bound by the van der Waals interaction. We applied STM/STS study on newly cleaved surfaces of $Fe_{1+\delta}Te$. The typical atomic topography is shown in Fig. 1a. The tetragonal lattice of terminal tellurium atoms form a stripe structure along *b*-axis, which is the result of the bi-collinear antiferromagnetic (AFM) order of this material[12,23,24]. The lattice parameter along *a*-axis orientation is about 3.7 Å, and 3.6 Å along *b*-axis orientation. The impurity image induced by the excess iron atoms can be viewed in this field of view (FOV) as bright spots. Figure 1b represents a typical STS spectrum of $Fe_{1+\delta}Te$ covering a voltage range of ±100 mV at T=1.4 K. The STS spectrum is V-shaped with an obvious dip structure exactly at the Fermi level. The scale of the DOS at high energy appears to be many times larger than that at Fermi level. The STS spectrum with a smaller energy range is shown in Fig. 1c. One can find that the minimum value of the STS spectrum is located exactly at the Fermi level. Some fine structures can be viewed at several millivolts with asymmetric distance to the zero-bias, which may have correlations with the charge density wave (CDW, e.g., the stripe structure) or the AFM order of this material[24]. Such a V-shaped STS spectrum is hard to be explained by the band edge effects because the dip is tightly pinned to Fermi level. Similar effect has been observed in previous experiment although the dip is not as sharp as our data shown here[12,24]. Theoretically it is assumed that the $Fe_{1+\delta}Te$ is a bad metal and is in the vicinity of an orbital-selective Mott transition as a result of intrinsic strong electronic correlation effect[10]. A general idea about the strong correlation effect is that the DOS near the Fermi level is supressed and transferred to the higher energy region[25]. The V-shaped STS spectrum on the surface of $Fe_{1+\delta}Te$ is thus the signature of strong correlation effect[11] although a quantitative description is still lacking.

The $Bi_2Te_3$ thin films are grown in-situ on $Fe_{1+\delta}Te$ substrates by the method of MBE. Although there is a lattice mismatch between these two materials, the films can still be grown very well due to the van der Waals interaction[19]. Figure 2a represents the topography of the high quality as-grown $Bi_2Te_3$ thin film. Figure 2b shows the height profile of the film along the red line in Figure 2a and a step of height of about 0.97 nm is clearly seen which corresponds to the thickness of one quintuple layer (QL) of $Bi_2Te_3$. The hexagonal lattice structure of $Bi_2Te_3$ can be viewed in Fig. 2c. The lattice parameters along *a*, *b*, *c* orientations are about 4.1, 4.3, and 4.2 Å, respectively. The slight lattice distortion may be caused by the modulation of the $Fe_{1+\delta}Te$ substrate. We applied the STS measurements on the surface of $Bi_2Te_3$ thin film with a thickness of 6QLs at the temperature of 1.5 K. The tunnelling spectrum is shown in the main panel of Fig. 2d as a solid blue line. Some distinguished features can be observed easily, such as a kink at the energy of about -275 mV labelled by a black arrow and a dip at the Fermi level. The linear part between -200 mV to -50 mV is dominated by the DOS of the topological surface states[18,26]. The insert in Fig. 2d represents a zoom-in DOS spectrum near Fermi level with a much higher energy resolution. The dip structure at the Fermi level can be viewed clearly covering the voltage range of about ±50 mV. One can easily see that the bottom of the dip locates at the Fermi level, the same as shown by the STS spectrum on the surface of $Fe_{1+\delta}Te$, as shown in Fig. 1b. To verify whether such a dip structure is caused by the effect from the $Fe_{1+\delta}Te$ substrate or just an intrinsic character of $Bi_2Te_3$ itself, we show the STS spectrum measured on $Bi_2Te_3$ thin films with the thickness of 7QLs on Si (7×7) substrates in the main panel of Fig. 2d plotted by the red solid line. This spectrum behaves in the same way as the one on $Bi_2Te_3/Fe_{1+\delta}Te$ except in the region near the Fermi energy, and the V-shaped dip near the Fermi level is totally absent. Both the spectra host an obvious knee structure as labelled by arrows. The energy value of the knee structure shifts because these two heterostructures have different extent of the charge transfer between films and substrates. A tiny kink appearing near Fermi level instead of the huge suppression of DOS near Fermi level is observed on $Bi_2Te_3$/Si, this may be attributed to the band edge effect of $Bi_2Te_3$. In previous study[18,21,22], it has been shown that the spectrum measured on the $Bi_2Te_3$ thin film grown on Si represents the intrinsic characteristics of $Bi_2Te_3$. It is thus tempting to conclude that the zero-bias dip structure of the STS spectrum on $Bi_2Te_3/Fe_{1+\delta}Te$ heterostructure is not an intrinsic character for $Bi_2Te_3$ but from the $Fe_{1+\delta}Te$ substrate. And we suggest that the strong electronic correlation effect inherited from $Fe_{1+\delta}Te$ substrate is responsible for this suppression of DOS near Fermi level with an energy range of about ±50 meV.

As a 3D topological insulator, $Bi_2Te_3$ hosts linearly dispersed SSs existing within the bulk band gap as shown in the schematic plot of band structure in Fig. 3d. The SSs behave as 2D electron gas on the surface of $Bi_2Te_3$. According to earlier studies, when the thickness of $Bi_2Te_3$ film exceeds 4QLs, the size effect (e.g., the interference between the upper and bottom surfaces of the $Bi_2Te_3$ film) is negligible and the band structure of the film resembles that of the bulk material[27]. In this case, the topological SSs are independent from the direct coupling of the two opposite surfaces[21,27-29]. When a magnetic field is applied perpendicular to the planes of $Bi_2Te_3$, the electrons of SSs fall into discrete Landau levels (LLs) [30]. The momentum and energy is quantized at the $n^{th}$ LL and the discrete energy scale $E_n$ is proportional to $\sqrt{n}$. The oscillation of DOS on the STS spectrum at LLs can therefore be used to determine the SSs band dispersion and the location of the Dirac point ($E_{DP}$ at $n$ = 0).

We conducted STS measurements on the 7QLs $Bi_2Te_3/Fe_{1+\delta}Te$ film at temperature of 1.5K with a 10T external magnetic field, as shown in Fig. 3a. The LLs can be viewed in the form of oscillations of DOS, which occurs on one single STS spectrum with the coexistence of the zero-bias dip structure at the Fermi level. To make the effect more visible, we normalized the STS spectrum by dividing the background. The distinct Landau quantization peaks are obtained as presented in Fig. 3b. The background is obtained by averaging the neighbouring 50 data points. The LLs indexes $n$ are then confirmed and labelled in Figs. 3a and 3b. The plot of $E_n$ (Landau level energy got from Fig. 3b) against $\sqrt{n}$ is shown in Fig. 3c according to the linear relationship between them. The typical linear behaviour demonstrates the existence of Dirac cone which represents the unbroken surface state. We then fitted the experimental data using the equation: $E_n = E_{DP} + v_F\sqrt{2e\hbar nB}$, and the obtained Fermi velocity $v_F$ of the Dirac electrons is $(4.53 \pm 0.02) \times 10^5$ (m/s) which coincides well with the value reported earlier[30]. The Dirac point energy is fitted to be -277.2$\pm$0.9 meV. According to the ARPES study on Bi$_2$Te$_3$[17], the energy difference ($\Delta E$) between the Dirac point ($E_{DP}$) and the bottom of the bulk conduction band ($E_{BCB}$) is about 295 meV as illustrated in Fig. 3d. We demonstrate that for this 7QLs film, $E_{BCB}$ is about 18 meV which is above the Fermi level. From above discussion, it is safe to conclude that the Fermi level resides in the bulk gap region, the DOS supressed near the Fermi level is closely related to the topologically protected surface states, which makes the phenomenon more peculiar and interesting.

In order to characterize the properties of the dip structure and its evolution as a function of thickness, we conducted STS studies on Bi$_2$Te$_3$ films with different thicknesses on Fe$_{1+\delta}$Te. Typical spectra obtained on 6QLs, 7QLs, and 8QLs films are plotted in Fig. 4a. Besides the zero-bias dip, these three spectra share the similar feature in general which results from the intrinsic band structure of Bi$_2$Te$_3$[14,18,26]. Because of the band bending effect, the spectra character from Bi$_2$Te$_3$ on 6QLs and 8QLs films have energy shifts with the value of about -32 meV and 17meV relative to the one on 7QLs films, as determined from the sharp steps of the knee structure. The energy values of the Dirac point and the bottom of the conduction band for the 6QLs and 8QLs films have been determined with the reference to the 7QLs film and the constant value of $\Delta E$ (295meV). The resultant $E_{BCB}$ positions marked as dashed lines are all near the Fermi level. In addition, all the three spectra exhibit the dip feature originated from Fe$_{1+\delta}$Te also near the Fermi level. Figure 4b represents spectra in a smaller voltage range with a higher energy resolution. The dip structures are clearly resolved for all three samples. The suppression of DOS at the Fermi level seems to get weaker with the increase of the thickness of $Bi_2Te_3$, while the dip position is pinned at the Fermi level robustly regardless of the shift of the band structure from $Bi_2Te_3$ among films with different thickness. This confirms that the dip structure on STS spectrum has nothing to do with the intrinsic band structure of $Bi_2Te_3$. The dip structure is not symmetric relative to Fermi level even the monotonically increasing background are deducted. The determined locations of the bottoms of bulk conduction bands are indicated by vertical dashed lines. One can see that, for the 6QLs film, the Fermi level locates in the bulk conduction band very close to the band edge; for the 7QLs and 8QLs films, the Fermi levels are already located in the bulk band gap region. As a consequence, for the 7QLs and 8QLs films, the DOS effected by the $Fe_{1+\delta}Te$ substrate is contributed by the topological surface states. Therefore, these $Bi_2Te_3/Fe_{1+\delta}Te$ heterostructures have merged the Dirac electrons and the effects inherited from $Fe_{1+\delta}Te$ which is a bad metal with strong correlation effect.

The STM/STS measurement normally detects the electronic properties at the top surface with a depth at most of about 10 Å. The zero-bias dip structure observed on surface of $Bi_2Te_3/Fe_{1+\delta}Te$ heterostructure originated from the $Fe_{1+\delta}Te$ substrate is quite puzzling, because the electronic states of $Fe_{1+\delta}Te$ can't be detected directly though the $Bi_2Te_3$ films with the thickness of several nanometers. There are several possible interpretations that may be applied to the zero-bias dip structure of $Bi_2Te_3$ films, such as size effect, lattice distortion, even interface induced superconductivity. Size effect can generate quantum well states in ultrathin films and thus change the distribution of intrinsic DOS. The direct SSs coupling destroys the Dirac cone and a hybridization gap opens at the Dirac point[28,31,32]. However, according to the previous studies[27,28], such a size effect is negligible for the 6QLs films and even thicker ones in our experiment, and has nothing to do with the observed novel phenomenon. An alternative picture would be the large lattice distortion which may also change the band structure of the film. However, if the band inversion has been destroyed by such distortion, the SSs will no longer exist[1-3]. There is negligible lattice distortion from our experiments, and the SSs exhibit very well and the Dirac point locates at the predicted energy value, therefore it is unlikely that the lattice distortion should be responsible for the suppression of the DOS near Fermi level. We also notice that interface superconductivity has been detected in $Bi_2Te_3$/FeTe-film heterostructure[33], but we observed neither the superconducting gap nor the superconducting resistive transition. The dip structure in the STS spectrum on 6QLs film covers a voltage range of about ±50 meV with asymmetric features relative to the Fermi level, and does not get weaker with the external magnetic field as high as 10 T. These characters have ruled out the effect of interface superconductivity or local pair.

By comparing the STS spectra measured on $Bi_2Te_3/Fe_{1+\delta}Te$ and $Fe_{1+\delta}Te$, we can conclude that the zero-bias dip structure observed on the surface of $Bi_2Te_3/Fe_{1+\delta}Te$ can only be understood as the result of strong electronic correlation effect[34] transferred from $Fe_{1+\delta}Te$ to the surface through the topological film $Bi_2Te_3$. As a general description of the behaviours of electrons, correlation effect results from several mechanisms and causes many phenomena[34]. The normal metal/Mott insulator heterostructures have been studied extensively by theorists[35-37]. In such a structure, the indication related to correlation effect, such as the antiferromagnetic order, should weaken rapidly in the interior of a metal, and become negligible after few metallic atom layers (several angstroms) adjacent to the interface[35-37]. However, in our $Bi_2Te_3/Fe_{1+\delta}Te$ heterostructures, considering the dip structure as the indication of the correlation effect, it stays robust even penetrating 8QLs films (about 8 nm, ten times larger than the theoretical penetrating depth referred above). How is the correlation effect robustly delivered from bottom to the surface of the TI film is mysterious, which challenges the possible ordinary understandings so far.

Concerning the combination of topological insulator and strong correlation effect, we suggest that the underlying physics is related to the peculiar SSs of the TI films. The Dirac electrons are massless and spin-momentum locked. Their abnormal behaviour deviates from the principles obeyed by normal electrons[1,2]. On the other hand, although the direct coupling of SSs no longer exists in the films, the Dirac electrons on the bottom can still tunnel to the surface by indirect coupling of SSs[38]. We believe that as the spin-polarized carriers, the Dirac electrons help to deliver the strong correlation effect from one side to another. The detailed mechanism is still unclear and needs to be resolved in the future study. This will stimulate a wide exploration on the similar heterostructures concerning abnormal features of the surface Dirac electrons.

In summary, we detected the merging effect of the Dirac electrons and strong correlation effect by constructing $Bi_2Te_3/Fe_{1+\delta}Te$ heterostructures. A pronounced zero-bias dip structure of DOS at Fermi level inherited from the $Fe_{1+\delta}Te$ substrate is transferred to the top layer of $Bi_2Te_3$ film, showing the coexistence of the Dirac electrons and the strong correlation effect. This abnormal behaviour of the strong correlation effect mixed with the surface Dirac electrons challenges many ordinary understandings. The underlying physics connected with this mysterious phenomenon needs to be explored in the future.

**Methods**

**Sample synthesis and film grown.** $Fe_{1+\delta}Te$ single crystals were grown by Bridgman method and cleaved in ultrahigh-vacuum chamber with pressure about $10^{-10}$ Torr. The Si (7×7) surface was obtained by annealing at about 1200℃. The films were grown in situ layer by layer under Te-rich condition in a Unisoku MBE chamber with pressure of about $5 \times 10^{-9}$ Torr. The growth rate is typically 1/10 QL per minute and determined by reflection high energy electron diffraction on Si (7×7) substrates.

**STM/STS measurements.** The STM studies were carried out with an ultrahigh-vacuum, low temperature and high magnetic field scanning probe microscope USM-1300 (Unisoku Co., Ltd.) with pressure better than $10^{-10}$ Torr. The STS spectra were measured by a lock-in amplifier with an ac modulation of 0.1 mV at 987.5 Hz to lower down the noise.

It should be noted that the films on $Fe_{1+\delta}Te$ substrates are grown with the theoretical thickness of 6QLs. A lot of spectra were measured on the films and the majority of them have the same feature and no energy shifts. This kind of spectra is determined to be the 6QLs $Bi_2Te_3$ film spectra. The terraces with the thickness of 7QLs and 8QLs are thus determined by the difference in height with the adjacent 6QLs terraces.

Surface Transport Channels in Bulk Insulating Bi$_2$Se$_3$ Thin Films. *Phys. Rev. Lett.* **113**, 026801 (2014).

**Acknowledgments**

We appreciate the kind help of Jinfeng Jia and xxx from Shanghai JiaoTong University in sharing the experience of MBE deposition. We also acknowledge the useful discussions with Qianghua Wang and Zhiping Yin. This work was supported by the Ministry of Science and Technology of China (973 projects: 2011CBA00102, 2012CB821403), NSF of China and PAPD.

**Author contributions**

The low-temperature STS measurements were finished by GD, ZYD, XY, DLF, HY, and HHW. The samples were prepared by EYW. The data analysis was done by GD and HY. HHW coordinated the whole work, GD, HY and HHW wrote the manuscript which was supplemented by others. All authors have discussed the results and the interpretation.

**Competing financial interests**

The authors declare that they have no competing financial interests.

[*] Correspondence and requests for materials should be addressed to hhwen@nju.edu.cn.

**Figure Legends**

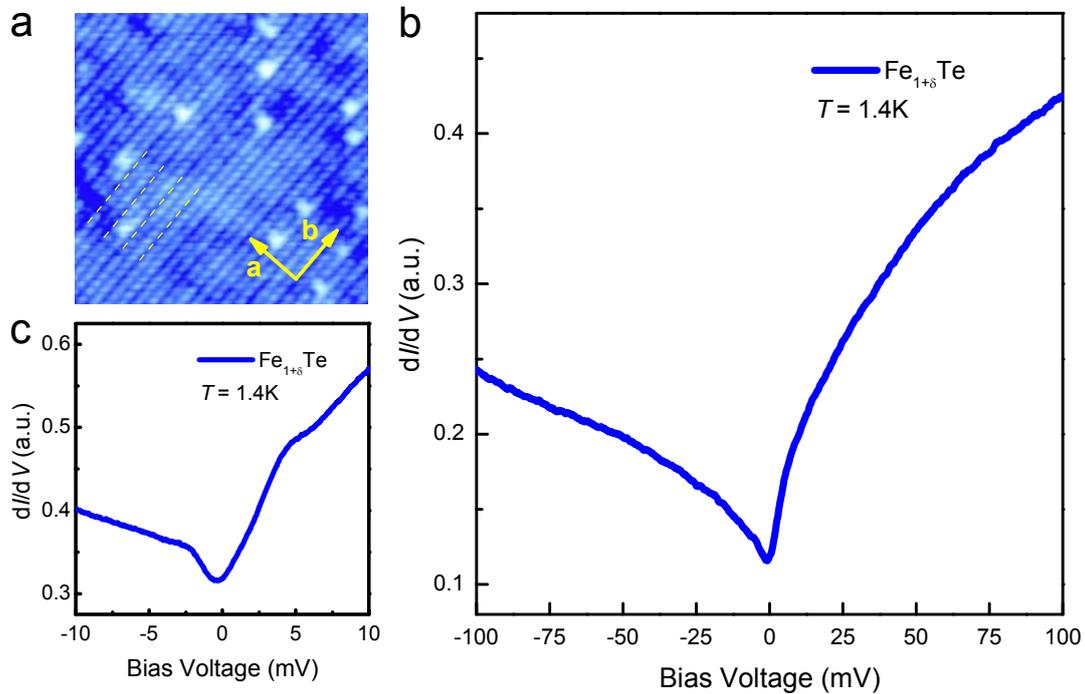

**Figure 1 | Topographic image and V-shaped STS spectrum of the Fe$_{1+\delta}$Te single crystal.**

**a,** Topographic image of Te lattice of Fe$_{1+\delta}$Te measured at 1.4 K (10nm ×10 nm, bias voltage $V_{bias}$ = 80 mV, tunnelling current $I_t$ = 0.1 nA). The centres of bright spots correspond to excess iron impurities. The yellow dashed lines in **a** indicate the stripe structure. **b,** The tunnelling spectrum measured on Fe$_{1+\delta}$Te, a "V" shape on the spectrum is clearly observed with the minimum pinned at the Fermi energy. **c,** The STS spectrum viewed in a smaller voltage range of ±10 mV.

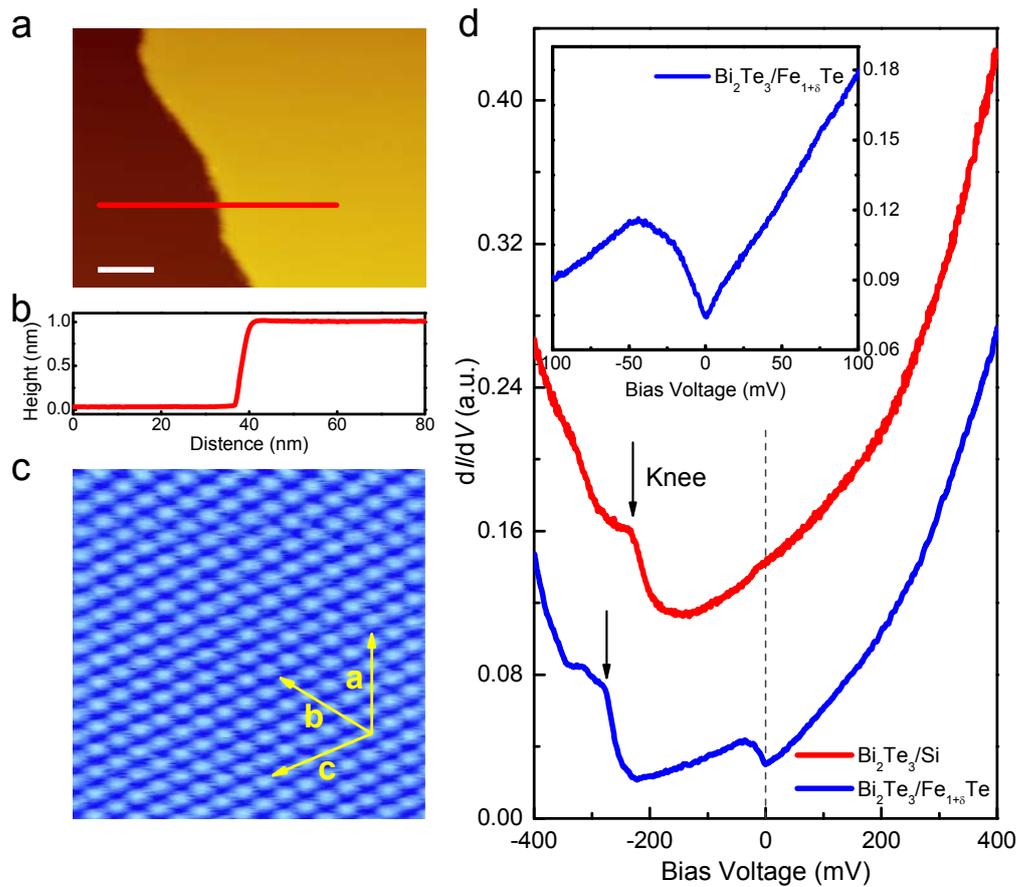

**Figure 2 | STM studies of the as grown Bi₂Te₃ thin films on Fe$_{1+\delta}$Te and silicon substrates. a,** Topographic image of the Bi₂Te₃ thin film grown on the Fe$_{1+\delta}$Te surface ($V_{bias}$ = 280 mV, $I_t$ = 14 pA). Scale bar, 20 nm. **b.** Line profile across a step structure along the red line in **a**, showing that the step is of 0.97 nm in height. **c,** Atomic image of the Te surface of Bi₂Te₃ film grown on Fe$_{1+\delta}$Te (6.6 nm × 6.6 nm, $V_{bias}$ = 50 mV, $I_t$ = 50 pA). **d,** Comparison of the STS spectra between the Bi₂Te₃ films (6QLs and 7QLs respectively) grown on Fe$_{1+\delta}$Te and Si(7×7) substrates. Two black arrows indicate the knee structure of the two spectra. Fermi level is labelled by a grey dashed line. The insert shows the STS spectrum of Bi₂Te₃ on Fe$_{1+\delta}$Te within a small voltage range.

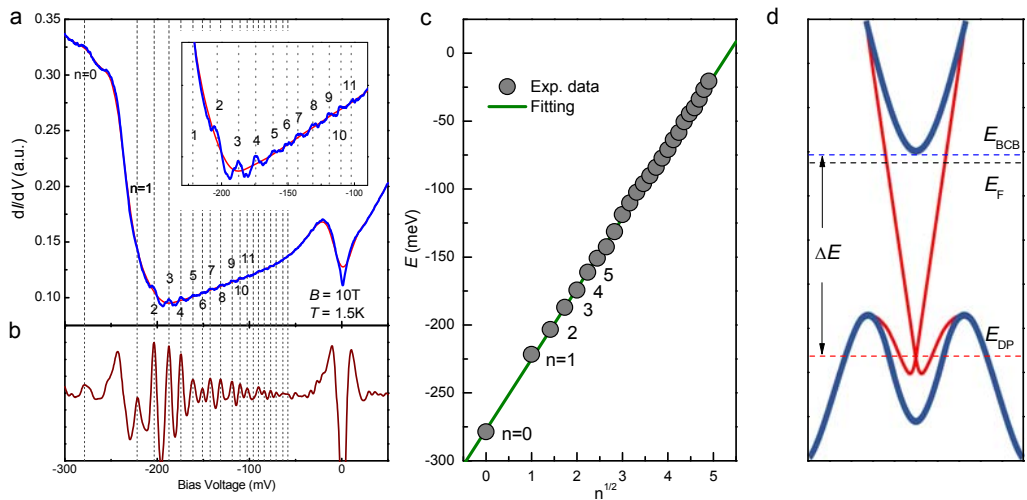

**Figure 3 | Landau levels observed on Bi$_2$Te$_3$/Fe$_{1+\delta}$Te. a,** d$I$/d$V$ spectrum measured at 10 T plotted in blue solid line in main panel with the background plotted in grey line. The background is obtained by smoothing of the experimental data. The insert represents the zoom-in of the spectrum in the main panel to illustrate the LLs. **b,** d$I$/d$V$ spectrum with the background being divided. The spectrum have been smoothed by averaging the neighbour data points to lower down the noise. The Landau levels are labelled by dark grey dashed vertical lines both in **a** and **b**. **c,** Peak positions of the Landau levels in **b** versus $\sqrt{n}$. The green line is the linear fitting to the experimental data. **d,** Illustration of the band structure of Bi$_2$Te$_3$. The topological SSs are indicated in red curves.

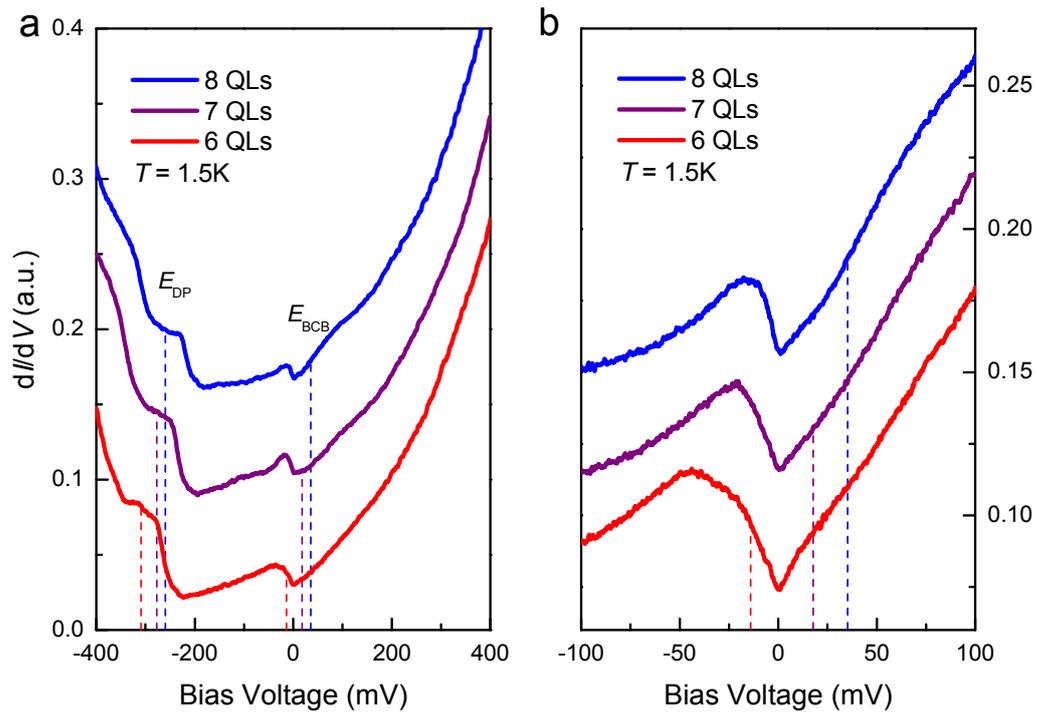

**Figure 4 | Properties of the dip structure near zero bias and evolution with thickness of $Bi_2Te_3$. a,** Tunnelling spectra measured on $Bi_2Te_3$ films grown on $Fe_{1+\delta}Te$ with different thickness. On each spectrum, the locations of $E_{DP}$, $E_{BCB}$ are indicated by dashed lines. **b,** Spectra with higher energy resolution in the voltage range of ±100 mV.